\documentclass[showkeys,showpacs]{revtex4}
\usepackage{amsmath}
\usepackage{amssymb}
\usepackage{graphicx}
\usepackage{color}

\begin{document}

\title{
Dynamical $F(R)$ gravities
}

\author{Vladimir Dzhunushaliev}
\email{vdzhunus@krsu.edu.kg} 
\affiliation{
Institute for Basic Research, 
Eurasian National University, 
Astana, 010008, Kazakhstan
}

\begin{abstract}
It is offered that $F(R)-$modified gravities can be considered as  nonperturbative quantum effects arising from Einstein gravity. It is assumed that nonperturbative quantum effects gives rise to the fact that the  connection becomes incompatible with the metric, the metric factors and the square of the connection in Einstein - Hilbert Lagrangian have nonperturbative additions. In the simplest approximation both additions can be considered as functions of one scalar field. The scalar field can be excluded from the Lagrangian obtaining $F(R)-$gravity. The essence of quantum correction to the affine connection as a torsion is discussed. 
\end{abstract}

\pacs{04.50.+h; 11.15.Tk}
\keywords{modified gravities; nonperturbative quantization effects}

\maketitle

\section{Introduction}

Current theoretical cosmology has two fundamental problems: what are inflation and dark energy ?  One possible resolution of these problems is the gravitational alternative (modified gravities or in other words: $F(R)-$gravities) for a unified description of the inflation and dark energy. From the author point of view at such resolution of these two problems erases  another problem: how the nature chooses from even uncountable sets of all $F(R)-$theories one of which is realized in the nature ?

Non-Einstein's gravities have a long history. They arise: by the perturbative quantization of 4D Einstein gravity \cite{Buchbinder} and by the consideration of quantum fields on the background of curved spacetime (for the textbook, see \cite{birrell}). For modern review on $F(R)-$gravities one can see \cite{Nojiri:2010wj}-\cite{Capozziello}. 

In this paper we consider a model where $F(R)-$modified gravities appear as \textcolor{blue}{\emph{nonperturbative}} quantum effects in Einstein gravity. It will be what we call as dynamical $F(R)-$gravities. As we see below nonperturbative quantization is either an operator differential equations \eqref{0-10} \eqref{0-20} or an infinite equations set \eqref{0-50}-\eqref{0-160}. If we could resolve either \eqref{0-10} \eqref{0-20} or \eqref{0-50}-\eqref{0-160} then we would have the necessary information on the quantum behavior of gravitational field. But in the consequence of extremely complicated problem for solving either \eqref{0-10} \eqref{0-20} or \eqref{0-50}-\eqref{0-160} we can not do it. Therefore we have to use an approximate approach. The main idea for such approximation is following: (a) in quantum gravity we will consider a connection $\Gamma$ and metric $g$ as independent variables; (b) by a nonperturbative quantization we assume that approximately 
$\left\langle Q \left| \hat \Gamma^2 \right| Q \right\rangle \approx \{ \}^2 + B(\phi)$ \footnote{where $\left.\left. \right| Q \right\rangle$ is a quantum state,  $\{ \}$ are Christoffel symbols and the function $B(\phi)$ describes the effects of the nonperturbative quantization of the affine connection $\Gamma$} and the metric factor 
$\left\langle Q \left| \widehat{\sqrt{-g}} \cdot \hat g^{\mu \nu} \right| Q \right\rangle$ in Einstein - Hilbert Lagrangian also has a quantum correction; (c) in the simplest approximation we assume that both quantum corrections are functions of one scalar function $\phi$; (d) considering the scalar field $\phi$ as an independent variable one can exclude the scalar field $\phi$ from the Einstein - Hilbert Lagrangian obtaining $F(R)-$modified theory. It means that $F(R)-$gravity appears after excluding nonperturbative quantum terms from averaged Einstein-Hilbert Lagrangian. 

In such approach there is a principal difference between standard and dynamical $F(R)-$gravities: For the first approach we should have either different functions $F(R)$ for an inflation era and a modern universe acceleration or one but very complicated function for the description both  regimes. In the second approach these different regimes are different quantum effects in one theory (quantum Einstein gravity).

\section{Nonperturbative quantization technique for gravity}

In 50th Heisenberg has offered the method for the nonperturbative quantization of a nonlinear spinor field \cite{heisenberg}. Following to Heisenberg,  nonperturbative operators of a quantum field can be calculated using corresponding field equation(s) for this theory. \textcolor{blue}{\emph{The main idea is that corresponding field equation(s) is(are) written for quantum field operators.}} 

According Heisenberg \cite{heisenberg} quantum operators of the metric 
$\hat g_{\mu \nu}$ obey to operator Einstein equations in the Palatini formalism 
\begin{eqnarray}
	\hat \Gamma^\rho_{\phantom{\rho} \mu \nu} -
	\frac{1}{2} \hat g^{\rho \sigma} \left(
		\frac{\partial \hat g_{\mu \sigma}}{\partial x^\nu} + 
		\frac{\partial \hat g_{\nu \sigma}}{\partial x^\mu} - 
		\frac{\partial \hat g_{\mu \nu}}{\partial x^\sigma}
	\right) &=& 0, 
\label{0-10}\\
	\hat R_{\mu \nu} - \frac{1}{2} \hat g_{\mu \nu} \hat R &=& 0
\label{0-20}
\end{eqnarray}
here for the simplicity we consider Einstein gravity without the matter; 
$\hat \Gamma^\rho_{\phantom{\rho} \mu \nu}$ is the operator of the connection; $\hat g_{\mu \nu}$ is the operator of the metric; $\hat R_{\mu \nu}$ is the operator of the Ricci tensor and defined in usual manner taking into that it is an operator 
\begin{eqnarray}
	\hat R_{\mu \nu} &=& \hat R^\rho_{\phantom{\rho} \mu \rho \nu}, 
\label{0-30}\\
	\hat R^\rho_{\phantom{\rho} \sigma \mu \nu} &=& 
	\frac{\partial \hat \Gamma^\rho_{\phantom{\rho} \sigma \nu}}
	{\partial x^\mu} - 
	\frac{\partial \hat \Gamma^\rho_{\phantom{\rho} \sigma \mu}}
	{\partial x^\nu} + 
	\hat \Gamma^\rho_{\phantom{\rho} \tau \mu} 
	\hat \Gamma^\tau_{\phantom{\tau} \sigma \nu} - 
	\hat \Gamma^\rho_{\phantom{\rho} \tau \nu} 
	\hat \Gamma^\tau_{\phantom{\tau} \sigma \mu} .
\label{0-40}
\end{eqnarray}
\textcolor{blue}{\emph{The nonperturbative quantization for Einstein gravity means that the quantum  operators $\hat \Gamma, \hat g$ obey the operator Einstein equations \eqref{0-10} \eqref{0-20}.}}

Now we have obtained an extremely complicated problem for solving operator Einstein equations \eqref{0-10} \eqref{0-20}. Heisenberg's approach for solving this problem is to write an infinite equations set for all Green functions
\begin{eqnarray}
	\left\langle Q \left| \hat \Gamma(x_1) 
	\cdot \text{ Eq. \eqref{0-10} }
	\right| Q \right\rangle &=& 0 ,
\label{0-50}\\
	\left\langle Q \left| \hat g(x_1) 
	\cdot \text{ Eq. \eqref{0-10} }
	\right| Q \right\rangle &=& 0 ,
\label{0-60}\\
	\left\langle Q \left| \hat \Gamma(x_1) 
	\cdot \text{ Eq. \eqref{0-20} }
	\right| Q \right\rangle &=& 0 ,
\label{0-70}\\
	\left\langle Q \left| \hat g(x_1) 
	\cdot \text{ Eq. \eqref{0-20} }
	\right| Q \right\rangle &=& 0 ,
\label{0-80}\\
	\left\langle Q \left| \hat \Gamma(x_1) \hat \Gamma(x_2) 
	\cdot \text{ Eq. \eqref{0-10} }
	\right| Q \right\rangle &=& 0 ,
\label{0-90}\\
	\left\langle Q \left| \hat g(x_1) \hat \Gamma(x_2) 
	\cdot \text{ Eq. \eqref{0-10} }
	\right| Q \right\rangle &=& 0 ,
\label{0-100}\\
	\left\langle Q \left| \hat g(x_1) \hat g(x_2) 
	\cdot \text{ Eq. \eqref{0-10} }
	\right| Q \right\rangle &=& 0 ,
\label{0-110}\\
	\left\langle Q \left| \hat \Gamma(x_1) \hat \Gamma(x_2) 
	\cdot \text{ Eq. \eqref{0-20} }
	\right| Q \right\rangle &=& 0 ,
\label{0-120}\\
	\left\langle Q \left| \hat g(x_1) \hat \Gamma(x_2) 
	\cdot \text{ Eq. \eqref{0-20} }
	\right| Q \right\rangle &=& 0 ,
\label{0-130}\\
	\left\langle Q \left| \hat g(x_1) \hat g(x_2) 
	\cdot \text{ Eq. \eqref{0-20} }
	\right| Q \right\rangle &=& 0 ,
\label{0-140}\\
	\cdots &=& 0	,
\label{0-150}\\
	\left\langle Q \left| 
	\text{ the product of $g$ and $\Gamma$ at different points $(x_1, 
	\cdots , x_n)$} 
	\cdot \text{ Eq. \eqref{0-10} or \eqref{0-20}}
	\right| Q \right\rangle &=& 0 
\label{0-160}
\end{eqnarray}
where $\left. \left|Q \right. \right\rangle$ is a quantum state, for example, it can be the quantum state for the inflation, the modern universe acceleration and so on. Below for the simplicity we will write $\left\langle \cdots \right\rangle$ instead of 
$\left\langle Q \left| \cdots \right| Q \right\rangle$. Schematically the first equation \eqref{0-50} has $\left\langle \Gamma^2 \right\rangle$ and 
$\left\langle \Gamma \cdot g \right\rangle$ terms; the second equation \eqref{0-60} has $g \cdot \Gamma$ and $g \cdot g^{\mu \nu} \cdot \partial g$; 
the third equation
\eqref{0-70} has $\left\langle \partial\Gamma \cdot \Gamma \right\rangle$ and 
$\left\langle \Gamma^3 \right\rangle$ terms and so on up to infinity. Thus all equations are linked and this is the main problem to solve such infinite equations set. Similar equations set one can find in statistical physics and turbulence theory. One can also remember the Heisenberg's matrix mechanics which is working with an infinite equations set for matrix elements. On the perturbative language Eq's \eqref{0-50}-\eqref{0-160} are Dyson - Schwinger equations but usually (for instance, in QED) they are written on the Feynman diagrams language. 

In Eq's \eqref{0-50}-\eqref{0-160} there are products like 
$\left. \hat \Gamma^n \right|_{x_i=x}$, $\left. \hat g^n \right|_{x_i=x}$ or $\left. \hat \Gamma^n \right|_{x_i=x} \cdot \left. \hat g^m \right|_{x_i=x}$. By using the perturbative approach (Feynman diagram technique) these products leads to singularities because the product of field operators (for free fields) in one point is poorly defined. We have to point out on the difference between such products by perturbative and nonperturbative quantization: for the nonperturbative quantization such singularities may be much softer or be absent in general \cite{heisenberg}. The matter is that in the perturbative approach we have moving particles (quanta) only. The communication between two spacetime points is carried by such quanta. Consequently the correlation function between these points (Green function) is not zero only if the interaction between points with quanta is possible. It means that corresponding Green function is not zero inside of a light cone and is zero outside of them. Such kind of functions have to be singular on the light cone (it is well known). But for the nonperturbative case it is not the case: for  self-interacting nonlinear fields may exist static configurations. For instance, it can be a nucleon, a glueball, a flux tube stretched between quark and antiquark and filled with a longitudinal chromoelectric field and so on.  In this case the correlation function (Green function) is not zero outside of the light cone. This is the reason why the Green function for the nonperturbative case is not so singular in the comparison with the perturbative case. 

Mathematically it can be explained in such a way: for free and interacting operators of quantum fields we have two \textcolor{blue}{\emph{different}}  algebras. For free fields the algebra is well known and it is described by canonical commutations relationships where commutators and anticommutators are distributions. But for interacting fields the algebra is unknown and above we bring forward arguments that defining relationships will be ordinary functions (not distributions). 

It is high probable that the equations set neither \eqref{0-10} \eqref{0-20}  nor \eqref{0-50}-\eqref{0-160} can not be solved analytically. 
One possible way to solve approximately these equations is following. We decompose $n-$th Green function 
\begin{equation}
	G_n = G(x_1, x_2 \cdots , x_n) = 
	\left\langle 
		\Gamma(x_1) \cdots g(x_m) \cdots 
	\right\rangle
\label{0-170}
\end{equation}
on the linear combination of the Green functions products of lower orders 
\begin{equation}
\begin{split}
	G_n(x_1, x_2 \cdots , x_n) \approx & G_{n-2}(x_3, x_4 \cdots , x_n)
	\left[ G_2(x_1, x_2) -C_2 \right] + 
\\
	&
	\left( \text{permutations of $x_1,x_2$ with
	$x_3, \cdots , x_n$} \right) + 
\\
	&	
	G_{n-3} \left( G_3 -C_3 \right) + \cdots 
\end{split}
\label{0-180}
\end{equation}
where $C_{2,3 \cdots}$ are constants. In such a way one can cut off the infinite equation set \eqref{0-50} - \eqref{0-160}. Such technique is applied for solving similar equations set in statistical physics and turbulence theory. 

Another way to solve approximately the infinite equation set \eqref{0-50} - \eqref{0-160} is choose some functional (for instance, the action or something like to gluon condensates in quantum chromodynamics \footnote{in quantum chromodynamics there exist various gluon condensates: 
$\mathrm{tr}\left(F_{\mu \nu} F^{\mu \nu} \right)$, 
$\mathrm{tr} \left( F_\mu^\nu F_\nu^\rho F_\rho^\mu \right)$ and so on, for the review on the gluon condensate one can see Ref. \cite{Zakharov:1999jj}. Using these condensates and integrating their one can obtain various functionals.}) and write down its average expression using corresponding Green functions. After that it is necessary to use some well-reasoned physical assumptions to express the highest order Green function through Green function of of lower order, see the decomposition \eqref{0-180}. Finally we will have a functional which can be used to obtain Euler - Lagrange field equations for Green functions. Below we will use such method for obtaining Einstein - Hilbert + scalar field action. 

Let us note that equations \eqref{0-50} - \eqref{0-160} are similar to: (a) a Bogolyubov chain of equations (hierarchy) for the one-particle, two-particle, etc., distribution functions of a classical statistical system; (b) Dyson - Schwinger equations for n-point Green functions in quantum field theory; (c) equations connecting correlation functions for velocities, pressure and density in a turbulent fluid. In all cases the solution methods are similar to discussed above: the decomposition of n-point corresponding functions on the product of lower function order. For us more interesting is the case (b): Dyson - Schwinger equations. In a perturbative quantum field theory these equations usually written on the language of Feynman diagram technique. But for us this approach is not inapplicable because the gravity is strongly self-interacting system. Heisenberg has applied nonperturbative technique for the quantization of a nonlinear spinor field \cite{heisenberg}. In Ref. \cite{Dzhunushaliev:2010qs} Heisenberg's technique was applied for the quantization in quantum chromodynamics and it was shown that such approach leads to the description of a gluon condensate in a glueball and in a flux tube filled with a longitudinal color electric field and stretched between quark - antiquark located at $\pm$infinities. 

The approaches from \cite{heisenberg, Dzhunushaliev:2010qs} are not applicable for the quantization of general relativity because the nonlinearities in gravity are extremely strong. In usual field theory the nonlinearities are in potential term but in gravity they are in kinetic term. Therefore we can not apply the nonperturbative quantization methods from usual quantum field theory to the quantization of gravity. 

\section{Nonperturbative calculations}

Excluding the hats $\widehat{(\cdots)}$ from equation \eqref{0-40} we obtain  the classical Riemann tensor for the affine connection $\Gamma^\rho_{\phantom{\rho} \mu \nu}$ as 
\begin{equation}
	\tilde R^\rho_{\phantom{\rho} \sigma \mu \nu} = 
	\frac{\partial \Gamma^\rho_{\phantom{\rho} \sigma \nu}}{\partial x^\mu} - 
	\frac{\partial \Gamma^\rho_{\phantom{\rho} \sigma \mu}}{\partial x^\nu} + 
	\Gamma^\rho_{\phantom{\rho} \tau \mu} 
	\Gamma^\tau_{\phantom{\tau} \sigma \nu} - 
	\Gamma^\rho_{\phantom{\rho} \tau \nu} 
	\Gamma^\tau_{\phantom{\tau} \sigma \mu} 
\label{1-10}
\end{equation}
where $\tilde{\phantom{R}}$ means that the scalar curvature $\tilde R$ is calculated for the affine connection $\Gamma$ which in general does not coincide with the Christoffel symbols $\{ \}$. Einstein - Hilbert Lagrangian is 
\begin{equation}
	\mathcal L = \sqrt{-g} g^{\mu \nu} 
	\tilde R^\rho_{\phantom{\rho} \mu \rho \nu} .
\label{1-20}
\end{equation}
In the classical approach varying with respect to independent variables the connection $\Gamma^\rho_{\phantom{\rho} \mu \nu}$ and the metric $g_{\mu \nu}$ leads to the fact that the connection becomes metric dependent 
\begin{equation}
	\Gamma^\rho_{\phantom{\rho} \mu \nu} = 
	\{^\rho_{\mu \nu}\}
\label{1-30}
\end{equation}
where $\{^\rho_{\mu \nu}\}$ are the Christoffel symbols. 

In general relativity we can use the metric and the connection as independent variables but using Palatini approach one can show that the connection is metric dependent. For quantum gravity again we work with the metric and the connection as with independent variables but now we assume that the Palatini relation \eqref{0-10} is true for the operators only. Since the Palatini relation is nonlinear according to the metric it is not true for the averaged quantities $\left\langle \Gamma \right\rangle$ and 
$\left\langle g_{\mu \nu} \right\rangle$. Consequently the affine connection will have some additional contributions in addition to Christoffel symbols:  torsion. 

Firstly we will consider the situation with the affine connection $\Gamma$. We suppose that the affine connection fluctuates about the Christoffel symbols. It means that the first natural assumption for the affine connection is following 
\begin{equation}
	\left\langle 
		\hat \Gamma^\rho_{\phantom{\rho} \mu \nu} 
	\right\rangle \approx 
	\{^\rho_{\mu \nu}\}
\label{1-40}
\end{equation}
but 
\begin{equation}
	\left\langle 
		\hat \Gamma^2
	\right\rangle \neq \{\}^2 .
\label{1-50}
\end{equation}
The second assumption is about the correlation between the affine connections in two points. We suppose that such correlation (Green) function has a classical term (arising from the Christoffel symbols) and some quantum correction: 
\begin{equation}
	\left\langle 
		\hat \Gamma^{\rho_1}_{\phantom{\rho_1} \mu_1 \nu_1} (x_1)
		\hat \Gamma^{\rho_2}_{\phantom{\rho_2} \mu_2 \nu_2} (x_2) 
	\right\rangle \approx 
	\{^{\rho_1}_{\mu_1 \nu_1} \} (x_1)
	\{^{\rho_2}_{\mu_2 \nu_2} \} (x_2) + 
	C^{\rho_1 \rho_2}_{\mu_1 \nu_1 \mu_2 \nu_2}
	\tilde B(x_1, x_2 )
\label{1-60}
\end{equation}
where $C^{\rho_1 \rho_2}_{\mu_1 \nu_1 \mu_2 \nu_2}$ is some constant; 
$\tilde B(x_1, x_2 )$ is some function approximately describing the effects arising from the nonperturbative quantization. 

Secondly we will consider the situation with the quantization of the metric. The metric is a part of Lagrangian as the factor $\sqrt{-g} g^{\mu \nu}$. A distinguishing feature of this term is that it has not any derivatives (let us remember that we consider the metric and the connection as independent variables). Therefore we assume that nonperturbative quantum corrections for the factor $\sqrt{-g} g^{\mu \nu}$ have following multiplicative form 
\begin{equation}
	\left\langle 
		\widehat{\sqrt{-g}} \; \hat g^{\mu \nu} 
	\right\rangle 
	\approx \sqrt{-g} \; g^{\mu \nu} \left( 1 + A \right)
\label{1-70}
\end{equation}
where $\widehat{\sqrt{-g}}, \hat g^{\mu \nu}$ are the operators of the measure $\sqrt{-g}$ and the contravariant metric $g^{\mu \nu}$. The term 
$A \gtrsim 1$ for a strong quantum regime and $A \ll 1$ for a weak quantum region.
 
Thirdly we assume that in the first approximation the factors $\widehat{\sqrt{-g}} \, \hat g^{\mu \nu}$ and $\hat{\tilde R}_{\mu \nu}$ do not correlate in the Lagrangian. Averaging Lagrangian \eqref{1-20} with the operators 
\begin{equation}
	\hat{\mathcal L} = \widehat{\sqrt{-g}} \, \hat g^{\mu \nu} 
	\hat{\tilde R}_{\mu \nu}
\label{1-75}
\end{equation}
under a quantum state 
$\left.\left. \right| Q \right\rangle$ and taking into account \eqref{1-60}  \eqref{1-70} we obtain following effective Lagrangian 
\begin{equation}
	\mathcal L_{eff} = 
	\left\langle \hat{\mathcal L} \right\rangle \approx 
	\left\langle \widehat{\sqrt{-g}} \, \hat g^{\mu \nu} \right\rangle 
	\left\langle 
		\hat{\tilde R}_{\mu \nu}
	\right\rangle \approx 
	\sqrt{-g} g^{\mu \nu} \left( 1 + A \right) \left[
		R + B
	\right] 
\label{1-80}
\end{equation}
where $R$ is the scalar curvature for the Christoffel symbols and $B$ is expressed through $C^{\rho_1 \rho_2}_{\mu_1 \nu_1 \mu_2 \nu_2}$ and 
$\tilde B$. 

The next assumption have to be done on the nonperturbative quantum corrections $A$ and $B$. We can not calculate their exactly. Most likely that by nonperturbative quantum calculations we have to do some well-founded physical assumptions of Green functions to cut off an infinite equations set for all Green functions. Bearing in mind this remark we assume that both functions $A$ and $B$ are function of one scalar field $\phi$: $A(\phi)$ and $B(\phi)$ (one scalar approximation). It leads to following effective Lagrangian 
\begin{equation}
	\mathcal L_{eff} = 
	\sqrt{-g} \left[ 1 + A(\phi) \right] \left[
		R + B(\phi)
	\right] .
\label{1-90}
\end{equation}
Lagrangian \eqref{1-90} is an intermediate Lagrangian on the way between  scalar-tensor and $F(R)$ gravities (for details one can see the review \cite{Nojiri:2010wj}). Varying corresponding action with respect to $\phi$ we will obtain the equation for $\phi$
\begin{equation}
	\left[ R + B(\phi) \right] A'(\phi) + 
	\left[ 1 + A(\phi) \right] B'(\phi) = 0 
\label{1-100}
\end{equation}
where $d(\cdots)/d\phi = (\cdots)'$. This equation could be solved with respect to $\phi = \phi(R)$. After that we can exclude function $\phi$ from the Lagrangian \eqref{1-90}
\begin{equation}
	\mathcal L_{eff} = 
	\left\langle \mathcal L \right\rangle \approx  
	\sqrt{-g} \left[ 1 + A(\phi(R)) \right]
	\left[
		R + B(\phi(R))
	\right] = \sqrt{-g} \, F(R)
\label{1-110}
\end{equation}
and obtain the $F(R)-$Lagrangian. 

\subsection{Reconstructing of $\Lambda$CDM-type cosmology}
\label{reconstructing}

In this subsection we would like to reconstruct $F(R)-$gravity having $\Lambda$CDM-type cosmology. In this case we will see what there should be the quantum corrections to have $\Lambda$CDM-type cosmology. 

Here we follow to Ref. \cite{Nojiri:2010wj}, section III C 1. The $F(R)$ action is 
\begin{equation}
\label{2-10}
	S_{F(R)}= \int d^4 x \sqrt{-g} \left[ 
		\frac{F(R)}{2\kappa^2} + \mathcal{L}_\mathrm{matter} 
	\right] .
\end{equation}
The action \eqref{2-10} can be equivalently rewritten in the form \eqref{1-90} as follows 
\begin{equation}
\label{2-20}
	S=\int d^4 x \sqrt{-g} \left\{
		\frac{1}{2\kappa^2} \left[ 
			P(\phi) R +  Q(\phi) 
		\right] + \mathcal{L}_\mathrm{matter}
	\right\} 
\end{equation}
where in our designation $1+A = P$ and $B=Q$ and the matter is added. To reconstruct the function $F(R)$ it is necessary to obtain the functions $P(R)$ and $Q(R)$. It can be done in the following way. The spatially flat FRW equations for $F(R)-$gravity are: 
\begin{eqnarray}
\label{2-30}
	-6 H^2 P(\phi) - Q(\phi) - 6H\frac{dP(\phi(t))}{dt} + 2\kappa^2
	\rho_\mathrm{matter} &=& 0 ,
\\
\label{2-40}
	\left(4\dot H + 6H^2\right)P(\phi) + Q(\phi) 
	+ 2\frac{d^2 P(\phi(t))}{dt^2}
	+ 4H\frac{d P(\phi(t))}{dt} + 2\kappa^2 p_\mathrm{matter} &=& 0 .
\end{eqnarray}
If we would like to reconstruct $F(R)-$modified gravity with the $\Lambda$CDM-type cosmology we have to consider \eqref{2-30} as the equation for the function $Q(\phi)$ and the difference \eqref{2-40}-\eqref{2-30} as the equation for the function $P(\phi)$ where Hubble parameter $H = \dot a/a$ is given as 
\begin{equation}
\label{2-50}
	H = \dot g, \quad 
	a = a_0 e^{g(t)} ,\quad
	g(t)=\frac{2}{3(1+w)}\ln \left\{
		\alpha \sinh \left[
			\frac{3(1+w)}{2l}\left(	t -	t_s \right)
		\right]
	\right\} 
\end{equation}
here, $t_s$ is a constant of the integration and
\begin{equation}
\label{2-60}
	\alpha^2\equiv \frac{1}{3}\kappa^2 l^2 \rho_0 a_0^{-3(1+w)} ,
\end{equation}
$w=p/\rho$ is the parameter of the equation of state; $l^2$ is the inverse cosmological constant and $\rho_\mathrm{matter}$ is given as 
\begin{equation}
\label{2-70}
	\rho_\mathrm{matter}=\rho_0 a^{-3(1+w_\mathrm{matter})}
\end{equation}
One can exclude the function $Q$ from Eq's \eqref{2-30} \eqref{2-40} obtaining inhomogeneous equation for the function $P$. Because this equation is a linear inhomogeneous equation, its general solution is given by the sum of the general solution that corresponds to the homogeneous equation
\begin{eqnarray}
\label{2-80}
	P = P_0 F(\tilde\alpha,\tilde\beta,\tilde\gamma; \phi)
	= P_0 \frac{\Gamma(\tilde\gamma)}{\Gamma(\tilde\alpha)
	\Gamma(\tilde\beta)} 
	\sum_{n=0}^\infty \frac{\Gamma(\tilde\alpha + n) \Gamma(\beta 
	+ n)}{\Gamma(\tilde\gamma + n)} \frac{\phi^n}{n!} ,
\\
	\tilde\gamma = 4 + \frac{1}{3(1+w)},\
	\tilde\alpha + \tilde\beta + 1 = 6 + \frac{1}{3(1+w)},\
	\tilde\alpha \tilde\beta = - \frac{1}{3(1+w)}
\label{2-90}
\end{eqnarray}
and of the special solution

\begin{equation}
	P=P_0(-\phi)^{-2/3} ,\quad P_0 
	= \frac{\eta}{\frac{10}{9} - \frac{2\left(\tilde\alpha + \tilde\beta 
	+ 1\right)}{3} + \tilde\alpha\tilde\beta} = - \frac{9\eta}{25}
\label{2-100}
\end{equation}
where $\Gamma$ is the $\Gamma-$function and $Q$ can be determined from equation \eqref{2-30}. 

In the limit of $\phi \to + \infty$, one can arrive at 
$P(\phi)R + Q(\phi) \to P_0 R - 6P_0/l^2$. Identifying $P_0=1/2\kappa^2$ and $\Lambda = 6/l^2$, the Einstein theory with cosmological constant $\Lambda$ can be reproduced.

Finally, having $P$ and $Q$ and recalling our designations $1+A = P$ and $B=Q$ we reconstruct what kind of quantum correlations should be to have $\Lambda$CDM-type cosmology. 

\subsection{The case $A,B =$ const}
\label{constant}

Here we would like to discuss separately the simplest case when above mentioned nonperturbative quantum corrections $A,B$ have the simplest possible form: they are constants $A,B =$ const. In this case the Einstein constant $\kappa$ is redefined in the following way 
\begin{equation}
	\frac{1}{2 \kappa^2_r} = \frac{1+A}{2 \kappa^2}
\label{3-10}
\end{equation}
and the cosmological constant $\Lambda$ is defined as 
\begin{equation}
	- 2 \Lambda = B. 
\label{3-20}
\end{equation}
Thus, if the quantum corrections for the connection and metric are constants then it leads to the $\Lambda$CDM-type cosmology directly. 

Now we would like to discuss in more details the essence of the quantum corrections of the affine connection. Let us remember that a torsion $Q^\mu_{\rho \sigma}$ is introduced in following way 
\begin{equation}
	\Gamma^\mu_{\rho \sigma} = \left\{ ^\mu_{\rho \sigma}\right\} + 
	Q^\mu_{\rho \sigma}
\label{3-30}
\end{equation}
where $\left\{ ^\mu_{\rho \sigma}\right\} = \Gamma^\mu_{\{ \rho \sigma \}}$ is the symmetrical part of the affine connection and the torsion 
$Q^\mu_{\rho \sigma} = \Gamma^\mu_{[ \rho \sigma ]}$ is the skew symmetrical part of the affine connection $\Gamma^\mu_{\rho \sigma}$. Then 
\begin{equation}
	\Lambda \approx g^{\mu \nu} \left\langle 
		\hat Q^\rho_{\tau \rho} \hat Q^\tau_{\mu \nu} - 
		\hat Q^\rho_{\tau \nu} \hat Q^\tau_{\mu \rho} 
	\right\rangle \equiv \left\langle \hat Q^2 \right\rangle 
\label{3-40}
\end{equation}
where $\hat Q^\mu_{\rho \sigma}$ is the torsion operator. Taking into account \eqref{3-40} we can estimate the variance of torsion (in the almost flat space $g^{\mu \nu} \approx 1$) as 
\begin{equation}
	\left\langle \hat Q^2 \right\rangle \approx \left| \Lambda \right| 
	\approx 10^{-57} \text{cm}^{-2}
	\quad \text{ with } 
	\left\langle \hat Q \right\rangle = 0.
\label{3-50}
\end{equation}
Thus in the presented approach the modern acceleration epoch is caused by an extremely small torsion effect. The torsion appears as a nonperturbative quantum correction to the Christoffel symbols. In principle the torsion effects can be measured. The main problem for testing this torsion approach to $\Lambda$ cosmological constant is that $\left\langle \hat Q \right\rangle = 0$ and $\left\langle \hat Q^2 \right\rangle$ is extremely small. 

\section{Conclusions}

The main result of this paper is that $F(R)-$modified gravities can be considered as some nonperturbative quantum effects arising from Einstein gravity. In the subsections \ref{reconstructing} and \ref{constant} it was shown what kind should be quantum corrections to have $\Lambda$CDM-type cosmology. For the calculations we have used some assumptions about nonperturbative quantum effects because up to now we do not have any quantum gravity theory. These assumptions are: (a) the connection and the metric are independent dynamic variables; (b) in the first approximation these quantum effects give rise to an additional terms for the squared connection and the metric factors in Einstein-Hilbert action. As the result one can obtain an effective Lagrangian with the Ricci scalar + scalar field approximately describing nonperturbative quantum effects. It allows us to exclude the scalar fields from the Lagrangian obtaining $F(R)-$modified gravity. 

The result of this paper can be formulated as follows. For the nonperturbative gravity quantization we have the infinite equations set \eqref{0-10} \eqref{0-20}. If we could solve these equations precisely and then exclude all quantum effects then we will obtain some very complicated gravity (even may be nonlocal) with 
$F(R), R^{\mu \nu}R_{\mu \nu}, R^{\mu \nu \rho \sigma}R_{\mu \nu \rho \sigma}$ and so on. That we have made here it is very simple approximation of such exclusion process. 

We have shown that for the modern epoch $\Lambda$-term can be considered as a torsion effect where the torsion appears as nonperturbative correction for affine connection. The symmetrical part of the affine connection is a classical part and the skew symmetrical part are quantum corrections. 

In the approach presented here both inflation era and modern acceleration epoch are quantum effects. For the first case it is not surprisingly because the Universe has started from a Planck scale. But why the modern acceleration is made from quantum effects ? In our opinion it is similar to a hypothesized flux tube in quantum chromodynamics. In quantum chromodynamics the SU(3) gauge field between quark and antiquark is confined into a flux tube stretched between quark and antiquark even they are located at $+\infty$ and $-\infty$. 

\section*{Acknowledgments}

I am grateful to the Research Group Linkage Programme of the Alexander von
Humboldt Foundation for the support of this research.

\end{document}